\documentstyle[aps,twocolumn]{revtex}
\input{psfig}
\begin{document}
\newcommand{\be}{\begin{equation}}
\newcommand{\ee}{\end{equation}}
\newcommand{\bq}{\begin{eqnarray}}
\newcommand{\eq}{\end{eqnarray}}
\draft
\title{Dynamical Mean Field Theory of the Bose-Hubbard Model} 
\author{Luigi Amico$^{*,\natural}$ and  Vittorio Penna$^{\dagger}$}
\address{$^{*}$Institut f\"ur Physik,Universit\"at Augsburg,
        Memminger Str. 6, D-86135 Augsburg, Germany.}
\address{$^{\natural}$ Istituto di Fisica, Facolt\'a di Ingegneria, Universit\`a
di Catania, viale A. Doria 6, I-95129 Catania, Italy.}
\address{
$^{\dagger}$ Dipartimento di Fisica, Politecnico di Torino and unit\`a INFM,
Corso Duca degli Abruzzi 24, I-10129 Torino, Italy.}
\maketitle
\begin{abstract}
Quantum dynamics of the Bose-Hubbard Model is investigated 
through a semiclassical hamiltonian picture provided by the
Time-Dependent Variational Principle method. 
The system is studied within a factorized slow/fast dynamics.
The semiclassical requantization procedure allows one to
account for the strong quantum nature of the system when $t/U\ll 1$.
The phase diagram is in good agreement with Quantum Monte Carlo
results and third order strong coupling perturbative expansion.    
\end{abstract}
\narrowtext
\bigskip
In recent years outgrowing interest has been devoted to many-body systems  
which can be modelled in terms of bosonic degrees of freedom.
Examples are granular superconductors, short-length superconductors and  
Josephson junction arrays~\cite{NATO}. The relevant physics of these systems
is captured by the Bose-Hubbard Model (BHM) which represents a boson gas of
identical charges hopping through a $D$ dimensional lattice. The boson
dynamics is described by the second quantized Hamiltonian
\be
H= \sum_{i} \left [U ( n_{i} -1) -\mu \right ] n_{i} -
{{t}\over{2}} \sum_{\langle i,j\rangle} \left (a^{\dagger}_{i} a_{j}+ 
a^{\dagger}_{j} a_{i} \right ),
\label{BHM}
\ee
where the operators $n_{i}\doteq  a^{\dagger}_{i} a_{i}$ count the number
of bosons at site $i$, while the annihilation and creation operators
$a_{i}$, $a^{\dagger}_{i}$ obey the canonical commutation relations 
$[a_{i},a^{\dagger}_{j}]=\delta_{ij}$. The  parameters of
Hamiltonian (\ref{BHM})
$U>0$, $t$, and $\mu$ correspond to the strength of the Coulomb
on-site repulsion, the hopping amplitude, and the chemical
potential, respectively. The rich structure of the BHM phase diagram
has been investigated by a number of 
theoretical methods, from mean field~\cite{FISHER},
variational~\cite{VARIATIONAL} and perturbative~\cite{FRMO} approaches,
to  Quantum Monte Carlo~\cite{QMONTE} technique. 

At T=0 and integer fillings, 
the system undergoes to a quantum phase transition,  
between a Mott Insulator (MI) and a Superfluid (SF)  phase. 
At $t=0$ the filling is fixed to  the integer $n$
that minimizes the onsite contribution of the
Hamiltonian (\ref{BHM}). 
In this limit, the existence of a finite energy $\mu=2 U$, 
required to add one  boson to the system, reflects the MI nature
of this phase, characterized by a vanishing compressibility 
(gapped particle-hole excitations).   
The MI  survives
(except for the degeneration points with $\mu/U = 2n$ )
when $t/U > 0$,
inside extended {\it lobes} in the  $t/U $, $\mu/U$ plane
attached to the intervals 
$I(n) = (\, 2(n-1)U \, , \, 2nU \,)$ of the $\mu/U$ axis~\cite{FISHER}.
Elsewhere, in the phase plane,
the system is superfluid and appears to be compressible since the
addition of a  particle free to hop through the lattice
brings up a reduction of the Coulomb repulsion.  
A crucial property has to be mentioned, at this point, concerning  
the MI-SF phase lobe boundary. At the transition  points the  appearence of
superfluidity is announced by the vanishing of the energy gap
between the states corresponding to $n$ ($n-1$) and $n+1$ ($n$) particles
(holes). Such a feature characterizes the whole frontier between the 
MI and the SF phase and will play a central role in the sequel. 

In spite of the great amount of work devoted to study the several aspects 
of the BHM, no investigation,
to the best of our knowledge, has been made on the
dynamics of boson degrees of freedoms. The attempt of relating the onset
of the macroscopic order leading to the MI and SF phases with the microscopic
behavior of the system motivates the dynamical approach to the BHM. This
should be useful as well to study both the emergence of vortex dynamics and
the transport properties. In particular,
in the present paper we shall show that the mean filed phase diagram of the
BHM results to be
quite improved if the {\it dynamics} of the SF order parameter is accounted 
for. 

Indeed it is possible to develop such a program,
although several features of Hamiltonian (\ref{BHM})
(nonlinearity, the many-body character and its quantum nature)
make the investigation of its dynamical behavior a hard problem.

The Time-Dependent Variational Principle (TDVP) method \cite{ZFG}
offers a quite general procedure for constructing an approximate macroscopic
wave function for many-body systems. Such a method, recently employed
for studying the fermionic Hubbard Model dynamics~\cite{MOPE}, 
is based on the idea of constraining the time evolution of the system's state
$|\Phi \rangle$ via the weaker form of the Schr\"odinger equation 
$\langle \Phi| (i\hbar \partial_{\tau} - H ) |\Phi \rangle = 0$.
Upon setting $|\Phi \rangle = exp({i {\cal S} / \hbar})|Z \rangle$ one obtains
\be
{\dot {\cal S}}
= i\hbar \langle Z| \partial_{\tau} |Z \rangle - 
\langle Z| H |Z \rangle
\label{azione}
\ee
where the {\it{trial macroscopic state}} $|Z\rangle$,
the basic ingredient of the method, must be structured so as to
contain as much information as possible on the microscopic dynamics.
The label $Z$ is thus identified with a vector of
microscopic parameters that are required to account for the dominating
physical processes at microscopic level and represent the effective dynamical
variables of the system. For this reason ${\dot {\cal S}}$ and
$\langle Z| H |Z \rangle$ can be properly seen as functions of $Z$ and
we denote them by ${\cal L}[Z]$ and ${\cal H}[Z]$, respectively.
The TDVP theory identifies the effective 
Lagrangian of the system with ${\cal L}[Z]$ and its semiclassical model
Hamiltonian with ${\cal H}[Z]$. The quantum dynamics is thus described
through a set of semiclassical hamiltonian equations obtained by
implementing the stationarity condition $\delta {\cal S} = 0$ on the
action ${\cal S}$ of ${\cal L}[Z]$.

A thorough analysis of the semiclassical equations related to $H$ is
interesting in itself due to their complexity. 
Here, we intend instead employ the TDVP method for finding the
ground-state configurations of the model and examining their dependence
from the microscopic parameters $t/U$ and $\mu /U$.
In other words, we aim to work out the phase diagram of
the BHM starting from the study of the semiclassical equations of motion.
Let us apply the TDVP method to the BHM model (\ref{BHM}). 
We write the state  $|\Phi \rangle$ as
$
|\Phi \rangle \equiv \exp{\left (i{\cal S} / \hbar \right ) } \otimes_{i} 
|z_{i}\rangle
$
once the trial macroscopic state $|Z \rangle$ is assumed to have the form
$|Z\rangle \equiv \otimes_{i} |z_{i}\rangle$. Here the states $|z_{i}\rangle$
are the Glauber coherent states~\cite{ZFG}
associated with the boson lowering
operators $a_{i}$, solutions of the equation
$a_{i}|z_{i}\rangle =z_{i} |z_{i}\rangle$ for each $i$.
The choice of $|Z \rangle$, of course, is suggested by the fact that
$H$ belongs to the enveloping algebra of the $N_s$-boson Weyl-Heisenberg algebra
$\{ {\bf I}, a_j , a_j^{\dagger}, n_j : j \in \Lambda \}$, 
$N_s$ being the number of sites of the lattice $\Lambda$.
In this case Eq. (\ref{azione}) becomes
\be
{{\cal L}}[Z] =
i\hbar \sum_i {\frac {1}{2}} 
({\bar z}_i {\dot z}_i - {\dot {\bar z}}_i z_i )
- {\cal H} (Z) \,\,,
\label{lagra}
\ee
where the semiclassical model Hamiltonian ${\cal H}(Z)=\langle Z|H|Z \rangle$
is easily shown to have the form
\be
{\cal H} = \sum_{i} (U |z_i|^2 -\mu ) |z_i|^2 -
{{t}\over{2}} \sum_{\langle i,j\rangle} 
\left ( {\bar z}_{i} z_{j} + {\bar z}_{j} z_{i} \right )\, .
\label{SBHM}
\ee
Lagrangian (\ref{lagra}) yields the equations of motion
\be
i \hbar \dot {z_i} =-\mu z_i +2 U z_i |z_i|^2 -
{{t}\over{2}}\, \sum_{j \in (i)} z_j \,\, ,
\label{SEE}
\ee
where $(i)$ indicates the set of the nearest neighbour sites around $i$ and
we have omitted the equation for ${\bar z}_{i}$ directly ensuing
from Eqs. (\ref{SEE}) via complex conjugation.
Notice that Eqs. (\ref{SEE}) can be obtained as well through the standard
formulas $i \hbar {\dot z_j } = \{ z_j , {\cal H} \}$, based on the canonical
Poisson Brackets $\{z_k ,{\bar z}_j \} = \delta_{kj} /i\hbar$ 
replacing commutators $[a_{i},a^{\dagger}_{j}] =\delta_{ij}$ in the
TDVP semiclassical scenery. This checks their hamiltonian character.
As expected, Eqs. (\ref{SEE}) are not integrable (integrability occurs 
only when either $U=0$ or when $t=0$) since the only known constant of motion,
a part from ${\cal H}$, is the semiclassical version
${\cal N} = \sum_i |z_i|^2$ of number operator $N = \sum_i n_i$.

We simplify the structure of Eqs. (\ref{SEE})
by separation of slow and fast dynamics, a procedure which is in some way
the analog, in a dynamical contest, of
the Mean Field Approximation (MFA) usually employed in Statistical
Mechanics. To this end we set, at each site, $z_i =\psi_i+\eta_i$ 
and assume that $\psi_i$ is a slow variable whereas $\eta_i$ is a fast
oscillating term describing the complex, high-frequency part of the dynamics
taking place on the microscopic interactions time scale (the hopping 
interaction, in this case). Then
$\psi_j = \langle z_j \rangle_{\tau}$ ($\langle \bullet \rangle_{\tau }$
denotes time average) when the time scale $\tau$ is larger than that
of the $\eta_j$'s. The onset of order in the system at the macroscopic scale
should reflect the dominating role of the $\psi_j$'s in the lattice dynamics.
Imposing the standard condition
$(z_i - \psi_i)({\bar z}_j - {\bar \psi}_j) = \eta_i {\bar \eta}_j \approx 0$
of the MFA procedure involves the dynamical mean field decoupling 
$z_i {\bar z}_j \approx \psi_i {\bar z}_j + 
{\bar \psi}_j z_i - \psi_i {\bar \psi}_j$ which implies, in turn, the
random phase approximation
$ \langle {\bar z}_j z_i \rangle_{\tau} \approx 
\langle {\bar z}_j \rangle_{\tau} \langle z_j \rangle_{\tau}$. The dynamical
scenery just depicted
together with an ergodic assumption leads thus naturally to defining
\be
\Psi \equiv {\frac {1}{N_s}} \sum_j \langle z_j \rangle_{\tau} 
\label{ORD}
\ee
as the macroscopic order paramenter revealing 
when order issues from  the lattice dynamics.

When our dynamical MFA is applied to Hamiltonian (\ref{SBHM}), and the
further assumption $\psi_{j} \equiv \psi_{i}$ for $j \in (i)$ 
is made (smoothing condition), the kinetic term modifies as follows
\be
{{t}\over {2}} \sum_{\langle i,j\rangle} \left ( {\bar z}_{i} z_{j}+ 
{\bar z}_{j} z_{i} \right ) \rightarrow
{{q t}\over{2}} \sum_{i}\left ({\bar z}_{i} \psi_{i}
+ \overline{\psi}_{i} z_{i} 
-|\psi_{i}|^2 \right ) \, ,
\label{SUB}
\ee
where $q$ denotes the number of nearest neighbours per site.
The resulting Hamiltonian reduces thus to a sum of on-site terms:
${\cal H}_{mf} = \sum_j {\cal H}_j$, where
\be
{\cal H}_j = U|z_{j}|^4 - \mu |z_{j}|^{2} - {{q t}\over {2}}
\left ( \overline{z}_{j} \psi_{j} +\overline{\psi}_{j}  z_{j} -
|\psi_{j}|^2 \right ) \, . 
\label{SH}
\ee
The hamiltonian equations ensuing from Eq. (\ref{SH})
\be
i \hbar \dot {z_{i}}=-\mu z_{i} +2 U z_{i} |z_{i}|^2 -
{\frac {qt}{2}} \psi_{i} \,,
\label{MFE}
\ee
bear memory of the off-site dynamics only through the on-site term
$\psi_{i}$. When compared with the exact ones, (\ref{SEE}), they imply
the relation $q \psi_{i} \approx \sum_{j \in (i)} z_j$ consistently
leading to an identity once time average is carried on and
the smoothing conditions are used.
It is important to notice that ${\cal N}=\sum_{i} |z_{i}|^2$ is no longer a
constant of motion while the mean field Hamiltonian
${\cal H}_{mf} = \sum_j {\cal H}_j$, that should approximate $\cal H$,
might represents a time-dependent total energy (which $\cal H$ is not)
since it depends on $\psi_{j}$'s. In order to recover such features,
it seems reasonable to introduce some restriction on the form of
$\psi_j(\tau)$, whose time behavior, so far, has not been specified at all.
We do this looking for solutions of Eqs. (\ref{MFE}) 
where $\theta_j$, $\chi_j$,
the phases of $z_j= |z_j| e^{i\theta_j}$, $ \psi_j= |\psi_j| e^{i\chi_j}$
respectively, are locked one to the other in such a way that
$\Delta \phi \doteq \theta_j - \chi_j = const$, the constant being
zero or $\pi$.
In view of Eqs. (\ref{MFE}), this entails
that any $|z_j|^2$ has a zero time derivative thereby restoring the particle
total number ${\cal N}$ to its proper role of time-independent quantity.
The same effect is obtained for any ${\cal H}_j$, and thus for
${\cal H}_{mf}$, only adding the further condition $|\psi_j|= const$.
Solutions of the form 
$z_j (\tau) = |z_j| e^{i\theta_j (\tau)} $ 
whose phase, due to Eq. (\ref{MFE}), obeys the equation 
\be
- \hbar |z_j| {\dot \theta}_j = (2U|z_j|^2 -\mu) |z_j| -{{qt}\over{2}} |\psi_j| \,\,,
\label{TETA}
\ee
successfully fulfill the conditions just stated.
Despite the elimination of any residual dynamical complexity we are able to
characterize  the MI  and some features of 
the SF phase via the phase dynamics of Eq. (\ref{TETA}).

We examine first the dynamics related to the MI. 
In this case, $\psi_i$
must have a zero time average along macroscopic time scales.
Such a behavior occurs when the uniform filling conditions
$n_i = n$, for all $i$ (we identify here number
operators $n_i$'s with their integer spectral values) is inserted in
(\ref{TETA}) by setting $|z_i|^2 = n$. Such a substitution is the
natural consequence of the requantization process~\cite{ZFG} of the
action-like variables $|z_i|^2$ (notice that
$\{ |z_i|^2 ,\theta_j \}= \delta_{ij}/ \hbar $) strongly requested from
the pure quantum character of the MI.
Eq. (\ref{TETA}) is thus solved by 
$\theta_j(\tau) = \lambda_\pm \tau / \hbar + \alpha_j$,
($\alpha_j$ is the initial condition) where
\be
\lambda_{\pm }\doteq {{1}\over{ \sqrt{n }}} 
\left ( U \delta \sqrt{n}
\pm {{qt}\over{2}}   |\psi_{i}| \right ) \; ,
\label{LAMBDA}
\ee
$\delta = \mu/U -2 n $
, and $\lambda_{-}$ ($\lambda_{+}$)  is related to the choice  
$\Delta \phi=\pi $ ($\Delta \phi=0$). 
Notice that the index $j$ does not label
$\lambda_\pm$ since the request $\langle z_j \rangle_{\tau} \equiv \psi_j(\tau)$
leads to $|\psi_j| = \sqrt n$ at each site.
In the present theory, the frequencies $\lambda_{\pm}$ 
play the role of time correlation lenght
governing the phase transition. 
Our theory gives  
$\lambda_{\pm} = U \sqrt{n} (\mu-\mu_c )$ for fixed  $t$ and
$\lambda_{\pm} =q |\psi_{i}| /2 (t-t_c)$ for  fixed $\mu$ 
($\mu_c$ and $t_c$ are the critical values of $\mu$ and $t$).
Defining the critical exponents $z$ and $\nu$ as in the ref.~\onlinecite{FISHER}, 
we argue that~\cite{COMMENT}
\begin{equation}
z \nu=1.
\label{EXP}
\end{equation}
By replacing  in the reduced Hamiltonian (\ref{SH}) 
the value of $|\psi_{i}|$ provided by
Eq. (\ref{LAMBDA}), the energy of the MI reads
\be
E_n (\mu , t;\lambda_\pm) = n \left[ U\delta - Un 
+{\frac{2}{qt}} (\lambda_\pm + U \delta )^2 \right] \; ,
\label{X1}
\ee    
where the subscript $n$
reminds us that the filling $n$ is accounted for. 
The oscillating behavior  of  
$\Psi =(e^{i \lambda_\pm \tau } /N_s ) \sum_j \psi_j e^{i \alpha_j} $, 
having a vanishing long time average, identifies the MI.
This, in fact, implies that the gauge symmetry breaking expected in the
SF phase cannot take place.
Notice that the ordinary (time independent) MFA
cannot describe the MI for $t>0$,
since the hopping term of the reduced Hamiltonian
is canceled  by the vanishing of the order parameter, $\psi=0$.
Within our scheme, instead, the condition
$\langle \Psi \rangle_\tau =0$ can be realized also for $\psi \neq 0$.

All dynamics disappears at the
degeneration  points $\mu/U=2n$, $t/U=0$, that represent the 
limiting points of the SF domain separating $I(n)$ from $I(n+1)$.
Within our theory, the fixed points of the 
equations of motion ($\lambda_\pm =0 $)
correspond to the SF system configurations. This is but the 
oversimplified version of the low frequency  dynamics 
expected in the SF phase. 
Considering such configurations (the trivial case $\dot z_j =0$ due to
$z_j = \psi_j =0$ is excluded) allows one to recast Eq. (\ref{MFE}) in the
form
\be
\psi_j = {\frac {2}{qt}} \, (2U|z_j|^2 - \mu ) z_j
\label{MBE} \,\,,
\ee
making $\psi_j$ a function of $z_j$. The energy associated 
to the hamiltonian ${\cal H}_j$, in turn, reduces to
$$
\epsilon (\mu , t, z_j) = |z_j|^2  
\left[ {\frac{2}{qt}}  (\mu - 2U|z_j|^2 )^2
+  (\mu - 3U|z_j|^2)    \right] \nonumber \, ,
$$
%
once ({\ref{MBE}) is inserted in (\ref{SH}). $\epsilon (\mu , t, z_j) $
is the on-site energy accounting for 
the absence of dynamics. 
The limit $\lambda_{\pm} \rightarrow 0$, in fact, shows that
$ E_n (\mu , t;\lambda_\pm)  \rightarrow  \epsilon (\mu , t, z_j)$ 
provided $|z_j|^2 = n$.
Its lowest value is found to be, via minimization,
$\epsilon(\zeta) = -U |\zeta|^4$, where the value
of $|z_j|^2$ corresponding to the minimum is $|\zeta|^2 = (\mu + 2t)/(2U)$.
It is worth noticing that inserting $|\zeta|$ in (\ref{MBE})
implies $\psi_j = z_j$ so that the minimum energy configuration
naturally satisfies the consistency condition on which our dynamical
MFA is based.

Let us consider, instead, how the MI characters reflect
on the macroscopic phase - viable to experimental observations -  ${\cal S}$.
Notice, first, that inserting Eqs. (\ref{SEE}) in the Lagrangian (\ref{lagra})
implies ${\dot {\cal S}} = U \sum_j |z_j|^4$ and that the same result
is found in the MFA scheme in that $\psi_j \approx z_j$. Then the exponential
factor of $| \Phi \rangle$ has a frequency ${\dot{\cal S}}$
which reduces to
\be
{\dot {\cal S}} =  U N_s n^2 \,\,\, ,
\label{ANGI}
\ee
when the system is a MI with $|z_j|^2 = n$ for each $j$. A transition leading
from the $n$-lobe to the $(n+1)$-lobe thus involves a change
of the phase frequency of $U N_s (2n+1)$, whereas in case of
a transition between any two superfluid states, where
${\dot {\cal S}_{mf}} =  U N_s |\Psi|^4$, since $z_j = \Psi$ for each $j$,
no quantization of the frequency variation appears, in that $\Psi$ takes
continuous values. 

Now, we employ the expression (\ref{X1}) for the on-site energy to
reconstruct the lobe-like structure of the phase diagram.
In the SF phase, the states with $n$ and $n+1$ (adding a particle), as 
well as the states with $n-1$ and $n$ (adding a hole) must be degenerate. 
The curves representing the $n$-lobe boundary are
identified  by implementing both gauge symmetry breaking 
through the limits $\lambda_\pm\rightarrow 0$ and  
vanishing of the energy gaps $E_n - E_{n \pm 1}$. In other
words we require
\begin{equation}
\lim_{\lambda_{+} \rightarrow 0}\left (E_n - E_{n+1} \right)
=0 \; \makebox{\hspace{0.5cm}} \left (\delta <0 \right ) \; ,
\label{REQ1}
\end{equation}
\begin{equation}
\lim_{\lambda_{-} \rightarrow 0}\left ( E_{n-1}-E_n \right )=0 \;
\makebox{\hspace{0.5cm}} \left (\delta >0 \right ) \;.
\label{REQ2}
\end{equation}
%
For solving Eqs.
(\ref{REQ1})-(\ref{REQ2})
we introduce the variables $\delta_{\pm} =\mu /U -2n+(1\pm 1)$.
By inserting $\delta_+ \ge 0$ ($\delta_-\le 0$) in eq. (\ref{REQ1})
((\ref{REQ2})), and  defining $r = qt/4U$, one gets the quadratic equations
$  \delta_{\pm}^2 + 2 r \delta_{\pm} - 2 r \, (2 n \mp 1) = 0 $
providing the curves
\be
{\frac {\mu_{\pm}} {U}} = 2(n-(1\pm 1))
- r \pm \left[ r^2 + 2 r\, (2 n \mp 1)  \right]^{1/2} \,\,.
\label{BB1}
\ee
The lower branch  $\mu_+ (t)$ and the upper $\mu_- (t)$
constitute the boundary encircling the $n$-th lobe.
We conclude by retrieving from Eq. (\ref{BB1}) the position of the farthest
point on the $n$-lobe boundary from the $\mu$-axis. By setting
$\mu_-(t) = \mu_+(t)$  one finds the lobe tip coordinates
\be
t_c = U/ qn
\label{TIP}
\ee
and $\mu(t_c)/U = 2n-1 -(1/2n)$. In the captions of Figs. (1)
and (2) the values of $t_{c}$ furnished by the present
approach, is  compared for the $D=1$ and $D=2$ cases
with QMC~\cite{QMONTE} and the Strong Coupling 
Perturbative Expansion (SCPE)~\cite{FRMO}.
\begin{figure}
\label{fig1}
\centerline{\psfig{figure=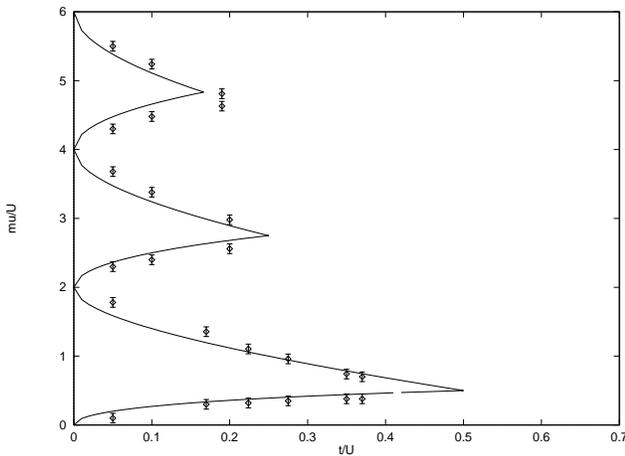,height=6cm}}
\caption{The Phase Diagram of the  BHM for $D=1$. The
error boxes are the QMC results of {\it Batrouni et al.}
Relatively to the first lobe ($n_{i}=1$), $(t_{c}/U)=0.5$.
QMC gives $(t_{c}/U)=0.43 \pm 0.002$ and the SCPE
$(t_{c}/U)=0.43$.
For ($n_{i}=3$), QMC and SCPE give  $(t_{c}/U)=0.2$ and 
$(t_{c}/U)=0.18$ respectively. Our theory gives $(t_{c}/U)=0.16$. }
\end{figure}

The dynamical approach we have developed 
appears to succeed in describing the quantum MI-SF phase
transition of BHM. 
The resulting phase diagram indeed exhibits an excellent
agreement with QMC simulations and SCPE results. 
This suggests that Eqs. (\ref{SEE}), here faced solely within the dynamical
MFA, deserve a systematic investigation by the methods of 
dynamical system theory. The dynamics they account for, in fact, should
describe not only zero temperature configurations but also excited states
involving density waves as well as vortices.
Moreover, a systematic analysis of Eqs. (\ref{SEE}) should be 
interesting both in relation to the dynamical scaling theory ~\cite{SACH}
and for calculating the dynamical correlation functions in the MI.
Work is in progress along these lines.

\acknowledgments

\vspace{-0.2cm}

We would like to thank G. Falci, R. Fazio, G. Giaquinta,   
M. Rasetti and C. Werner for useful discussions. The financial support of INFM (Italy) 
and of "Fondazione A. Della Riccia" is gratefully  acknowledged.
%

\vspace{-0.2cm}

\begin{figure}
\label{fig2}
\centerline{\psfig{figure=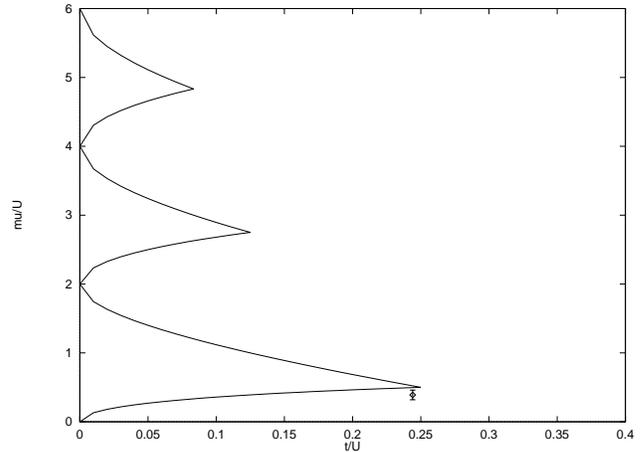,height=6cm}}
\caption{The Phase Diagram of the  BHM for $D=2$.
The error box indicates the QMC tricritical point
obtained by {\it Krauth and Trivedi}.
For $n_{i}=1$, $(t_{c}/U)=0.25$ while
QMC gives $(t_{c}/U)=0.244 \pm 0.002$ and SCPE provides $(t_{c}/U)=0.272$. 
}
\end{figure}

\end{document}